\documentclass[pdflatex,sn-mathphys-num]{sn-jnl}
\theoremstyle{thmstyleone}%
\theoremstyle{thmstyletwo}%

\theoremstyle{thmstylethree}%

\usepackage{graphicx}%
\usepackage{subcaption}%
\usepackage{multirow}%
\usepackage{amsmath,amssymb,amsfonts}%
\usepackage{amsthm}%
\usepackage{mathrsfs}%
\usepackage[title]{appendix}%
\usepackage{xcolor}%
\usepackage{textcomp}%
\usepackage{manyfoot}%
\usepackage{booktabs}%
\usepackage{algorithm}%
\usepackage{algorithmicx}%
\usepackage{algpseudocode}%
\usepackage{listings,comment}%

\usepackage[table]{xcolor}
\usepackage{csquotes}

\begin{document}

\title[]{Economic complexity at subnational level: \\A consistency analysis}


\author[1]{\fnm{Wenli} \sur{Du}}

\author[2]{\fnm{Andrea} \sur{Zaccaria}}

\affil[1]{\orgdiv{Modelling Engineering Risk and Complexity}, \orgname{Scuola Superiore Meridionale}, \city{Naples}, \postcode{80138}, \country{Italy}}

\affil[2]{\orgdiv{Istituto dei Sistemi Complessi}, \orgname{Consiglio Nazionale delle Ricerche}, \city{Rome}, \postcode{00185}, \country{Italy}}


\abstract{Several network‑based measures have been proposed to assess the economic complexity of countries. These measures have provided important insights into national economic development, and they are now widely applied at the subnational level as well. Here, we show that such applications lead to inconsistent results, in the sense that the estimated complexity of the same product appears to depend on methodological details such as the geographical scale of analysis. Building on these findings, we propose a measure of territorial economic complexity based on an exogenous and extensive computation. We show that these methodological choices yield estimates that are more consistent and more strongly aligned with standard economic indicators, such as GDP per capita and employment.}

\maketitle

\section{Introduction}\label{sec1}

Economic Complexity \cite{hidalgo2009building,balland2022reprint,neffke2024economic} is a field of study that aims to provide a holistic measure of the productive capabilities of large-scale economic systems, typically at the level of cities, regions, or countries. Specifically, economic complexity seeks to quantify how knowledge and skills (know-how) are accumulated within a collective, and how these capabilities are reflected in the economic activities observed across different regions. The economic complexity framework explains countries’ competitiveness in global markets using non-monetary and non-income-based indicators. A central question is whether a country’s productive capabilities can be inferred from the structure of its export basket \cite{hausmann2007export}. Rather than focusing on individual inputs such as capital, technology, or labor, the economic complexity approach begins with observed outcomes and examines economic development from the perspective of complex networks. Its core measurement logic is outcome-based: by analyzing the set of products an economy is able to produce, one can infer its underlying productive capabilities and thereby assess its overall level of development~\cite{bishop2019exploring}.

A substantial body of research has applied the economic complexity framework to provide empirically grounded and highly interpretable analyses of economic development at both national and regional levels. These studies offer valuable references and directional guidance for the formulation of differentiated development strategies \cite{hidalgo2021review,JRC138666}. At the aggregate level, higher economic complexity generally implies a more advanced and well-structured internal economic system, which is conducive to further economic growth ~\cite{hausmann2014atlas,sbardella2018role}. The ability of a country to produce a diverse range of products is a key indicator of the strength of its underlying productive capabilities \cite{tacchella2012fitness}. Empirical evidence shows that the economic complexity approach has strong predictive power for GDP growth ~\cite{castaneda2022growth,sbardella2018role}. For instance, it can predict future economic trajectories by identifying historically similar positions in the GDP per capita–fitness space ~\cite{cristelli2015heterogeneous,tacchella2018gdp}, thereby enabling assessments of long-term economic growth~\cite{pugliese2017poverty}. It also provides guidance for the selection of trade partners ~\cite{ren2024trading} and serves as an important decision-support tool for numerous international organizations~\cite{JRC138666,hidalgo2023policy}.

Throughout the development of economic complexity measurement methods, numerous approaches have been proposed, interpreted, and refined. These methods generally produce two key outputs: measures of the complexity of economies and measures of the complexity of products~\cite{hidalgo2021review,neffke2024economic}. PRODY~\cite{hausmann2007export} represents one of the earliest attempts to infer the productive capabilities required for a given product from the structure of countries’ export baskets. It is constructed based on the share of a specific product in a country’s total exports within the international trade network, combined with the exporting country’s GDP level. A widely used linear iterative method, the method of reflections~\cite{hidalgo2009building}, leads to the assessment of the Economic Complexity Index (ECI) of territories and the Product Complexity Index (PCI) of industries. The Economic Fitness and Complexity (EFC) method ~\cite{tacchella2012fitness} is a nonlinear iterative map to compute the capability content of countries (their "Fitness") and products (their "Complexity"). Both the founding principles and the outputs of these methods differ \cite{pietronero2017economic}, leading to methodological attempts to reconcile them ~\cite{sciarra2020reconciling}. Moreover, building on these frameworks, several extensions have been proposed. The Extensive Fitness \cite{tacchella2012fitness,patelli2022MS} gives greater weight to economies' size by applying a market-share weighting to the export matrix. The Exogenous Fitness \cite{operti2018dynamics} incorporates product-level information derived from the global export data to compute the economic complexity of subnational territories \cite{cattaruzzo2026specialization}. In addition, some approaches construct network structures based on domain-specific information, leading to economic complexity indicators relative to jobs and skills ~\cite{aufiero2024jobfitness} and to technological and scientific activities ~\cite{pugliese2016convergence,ivanova2017economic,balland2017geography,cimini2014scientific}. 


\begin{table}[t]
\caption{Sample literature on Economic Complexity at subnational level}
\label{tab:lit}
\centering
\renewcommand{\arraystretch}{1.2}
\setlength{\tabcolsep}{4pt}

\begin{tabular}{p{0.16\linewidth}p{0.2\linewidth}p{0.54\linewidth}}
\toprule
\rowcolor{blue!18}
\textbf{Authors} & \textbf{Territories} & \textbf{Results} \\
\midrule

\rowcolor{blue!10}
Chávez (2017)
& Mexico's states 
& Economic complexity can explain economic growth across Mexico's states: the higher a state's Economic Complexity Index (ECI), the higher its GDP per capita\\

\rowcolor{blue!5}
Chakraborty (2020)
& Japan’s prefecture
& ECI and fitness for the prefectures shows a high correlation with macroeconomic indicators, such as per-capita gross prefectural product and prefectural income per person \\

\rowcolor{blue!10}
Çınar (2022)
& Turkey’s regions
& Economic fitness positively affects regions’ overall labor productivity, and this positive effect mostly stems from the industrial sector \\

\rowcolor{blue!5}
Ferraz (2025)
& Brazil's municipalities 
& Both foreign direct investment (FDI) and absorptive capacity (AC) positively influence regional economic complexity  \\

\rowcolor{blue!10}
Freitas (2024)
& Brazilian microregions
& The distribution of high complexity industries in Brazil is very spatially concentrated, mainly in the microregions of São Paulo \\

\rowcolor{blue!5}
Fritz (2021)
& US metropolitan areas
& The relationship between economic complexity and income exhibits opposite patterns across different dimensions, and should therefore be interpreted with caution\\

\rowcolor{blue!10}
Gao (2018)
& China’s provinces
& ECI is positively associated with economic development and negatively associated with income inequality. In addition, both ECI and the Fitness demonstrate superior explanatory power for macroeconomic monetary indicators relative to conventional benchmark measures \\

\rowcolor{blue!5}
Herrera (2021)
& Brazil's states
& The states with the highest ECI are in the Southeast and South regions, and that in some states ECI declined or stagnated\\

\rowcolor{blue!10}
Mau (2019)
&  China’s regions
& On average, an increase in ECI is associated with faster per capita income growth \\

\rowcolor{blue!5}
Mealy (2019)
& U.K. local authorities and U.S. states
& Strong correlations exist between economic complexity, GDP, and income across international datasets, as well as within U.K. local authorities and U.S. states; however, no correlation is found between diversification and GDP per capita in the U.S. and the U.K. \\

\rowcolor{blue!10}
Poncet (2013)
& China’s cities
& The more complex the city’s productive structure, the faster its subsequent GDP per capita growth \\

\rowcolor{blue!5}
Reynolds (2018)
& Australia’s states and territories
& The ECI for each state and territory reflects relative economic complexity, with higher scores indicating more complex economies \\

\rowcolor{blue!10}
Roberto (2024)
& Italy’s provinces
& Higher economic complexity increases banks’ access to successful projects and their willingness to lend\\

\rowcolor{blue!5}
Sbardella (2017)
& US counties
& Wage inequality increases with growing economic Fitness \\

\rowcolor{blue!10}
Smolski (2024)
& Brazilian microregions
& Economic complexity is positively associated with GDP per capita in microregions, but its effects on growth vary across regions \\

\rowcolor{blue!5}
Teixeira (2022)
& Brazil’s federative units
& Economic complexity is positively associated with growth and productive transformation, though the effects are sometimes statistically weak \\

\rowcolor{blue!10}
Yeerong (2024)
& Thailand’s provinces 
& ECI is associated with economic growth after
surpassing a complexity threshold \\

\bottomrule
\end{tabular}
\end{table}

Beyond its initially well-developed applications at the international trade network level, the economic complexity framework has increasingly been extended to analyses within countries across different geographical and administrative scales, leading to studies using more fine-grained data at the subnational level. Usually, the authors apply one of the economic complexity algorithms to a matrix connecting subnational territories with economic activities, as traced by employment or exports, and compare the results with local macroeconomic quantities. A selection of papers is described in Table \ref{tab:lit}.\\
Cinar (2022)~\cite{cinar2022fitness} shows that, in Turkey’s regions, Economic Fitness positively affects overall labor productivity. Chavez (2017)~\cite{chavez2017complexity} finds, using the ECI method, that regarding Mexico’s states, higher levels of economic development are associated with higher ECI. More specifically, Mau (2019)~\cite{mau2019growth} demonstrates that, across China’s regions, an increase in ECI of approximately 0.10–0.15 standard deviations is associated with 1.3–1.8\% faster income growth per capita. Reynolds et al.(2018)~\cite{reynolds2018subnational} apply ECI for Australia’s states and territories. At the provincial level in Italy, Basile (2024)~\cite{basile2024credit} finds that higher local economic complexity increases banks’ access to successful projects and their willingness to lend, while Yeerong (2024)~\cite{yeerong2024ml} shows that ECI promotes inclusive and regionally distinctive economic development in Thailand’s provinces. Using both linear and nonlinear complexity measures, Teixeira (2022)~\cite{teixeira2022complexity} finds a positive relationship between economic growth and economic complexity across Brazil’s federative units, although this relationship is not always statistically significant. Herrera (2021)~\cite{herrera2021complexity} further highlights substantial heterogeneity in economic complexity across Brazilian states, noting that the highest ECI values are concentrated in the Southeast and South regions, while in some states, ECI has declined or stagnated. Gao (2018)~\cite{gao2018complexity} shows that both ECI and Fitness have stronger explanatory power for economic growth than conventional benchmark methods; additionally, ECI is positively associated with economic development but negatively associated with income inequality. Sbardella (2017)~\cite{sbardella2017inequality} finds, using U.S. county-level data, that Fitness is positively but nonlinearly related to wages, with deviations providing insights into regional development potential. Fritz (2021)~\cite{fritz2021complexity} shows that, for U.S. metropolitan areas, economic complexity measures (ECI and EFC) can capture economic performance; however, economic complexity Index (ECI) and income per capita may exhibit opposite patterns across different dimensions and should therefore be interpreted with caution. Accordingly, further research is needed to better assess the feasibility of using complexity as a predictor of future economic growth. In addition, Capoani (2024)~\cite{capoani2024diversification} highlights the importance of diversification in enhancing economic stability and resilience at the provincial level in Italy. By contrast, Mealy (2019)~\cite{mealy2019interpreting} finds no correlation between diversification and GDP per capita for the United States and the United Kingdom, although strong correlations are observed between economic complexity, GDP, and income for U.K. local authorities and U.S. states. Overall, these findings suggest that, at the subnational level, the use of different methods can lead to  different results even in similar research contexts.

Poncet (2013)~\cite{poncet2013complexity} shows, at the level of Chinese cities, that the more complex a city’s productive structure, the faster its subsequent GDP per capita growth. Qiao (2025)~\cite{qiao2025place}, using four-digit manufacturing industry data from China, finds that when cities diversify into more related and more complex industries, they tend to exhibit stronger path dependence. Balland (2020)~\cite{balland2020complex} finds, based on 353 metropolitan areas in the United States, that the spatial concentration of production activities increases with their level of complexity. Ruan (2025)~\cite{ruan2025green}, using panel data from 288 Chinese cities, provides evidence that the EFC method significantly promotes the development of green technology innovation (GTI). Chakraborty (2020)~\cite{chakraborty2020prefectures} finds that, at the prefectural level in Japan, both ECI and EFC are strongly correlated with macroeconomic indicators such as GDP per capita and income. Dong (2021)~\cite{dong2025green}, using three-digit city-level industry data, shows that cities providing higher levels of land subsidies are more likely to enter emerging industries with higher economic complexity (ECI). Freitas (2024)~\cite{freitas2024related} finds that industrial development in Brazilian microregions is neither sufficiently diversified nor balanced, with high-PCI industries primarily concentrated in São Paulo. Meanwhile, Smolski (2024)~\cite{smolski2024growth}, using an ECI-based dynamic economic complexity model, shows that economic complexity is positively associated with GDP per capita in Brazilian microregions, although its impact on GDP per capita growth differs across major regions.  Ferraz (2025)~\cite{ferraz2025fdi} examines the relationship between economic complexity, foreign direct investment, and absorptive capacity at the municipal level in Brazil.

In summary, the literature contains numerous contributions linking subnational economic complexity measures to macroeconomic outcomes. Most studies report a positive association, although there are notable exceptions. To the best of our knowledge, however, none of these contributions has carried out a systematic comparison across geographical scales, despite the fact that economic systems observed at different scales—such as countries and firms— exhibit distinct structural properties: country–product bipartite networks typically display nested patterns, whereas firm-level networks are more often characterized by modular structures \cite{laudati2023ecosystems}. Consistent with this concern, Diem (2024)~\cite{diem2024predictability} shows that using data at different levels of aggregation can lead to substantial differences in economic measurement outcomes. A further gap is the lack of a dedicated investigation of how results change when varying the granularity of product classifications.
Moreover, the output of economic complexity algorithms is intrinsically twofold: they produce an estimate of the complexity of territories and, simultaneously, an estimate of the complexity of products. Yet, while the former has been extensively scrutinized, the latter is almost always overlooked. In this paper, we therefore begin by analyzing product complexity, because these measures refer to the same products—independently of the country considered or the geographical scale—making them directly comparable across settings. This comparison reveals a pronounced lack of correlation between global (country-level) and local (subnational) product-complexity estimates, leading us to conclude that applying economic complexity algorithms at subnational scales yields inconsistent outputs.
Another recurrent feature of the existing literature is that, in the vast majority of cases, studies do not benchmark alternative economic complexity metrics against one another—implicitly assuming that they all extract the same information about the territorial presence of capabilities, as they should in principle. Here, we explicitly perform this comparison and show that these measures can be grouped into distinct families, whose outputs are correlated within the same family but largely uncorrelated across families. Finally, we use standard macroeconomic variables to identify which family of measures most reliably captures local economic development, and whether such performance remains stable across geographical scales.

\clearpage

\section{Results}\label{sec2}
\subsection{Inconsistency of subnational product complexity}
In the introduction, we showed that applying economic complexity metrics at the subnational level often leads to incoherent and contradictory results. The use of different metrics (which are intended to capture the same theoretical constructs), the heterogeneity of countries and spatial scales considered, as well as the use of different controls in regression analyses, make a comparative assessment of these findings challenging (see Table \ref{tab:lit}).
In this work, we conduct a parallel analysis of subnational data for three countries—Brazil, China, and Italy—with the aim of systematically comparing these metrics and identifying the sources of their discrepancies. Specifically, we construct a series of territory–product matrices in which we can vary both the geographical scale and the industrial classification’s level of product detail. The procedure adopted for building these matrices is described in detail in the Methods section.
From the territory–product matrices, we compute several economic complexity metrics: one associated with rows (territories) and one associated with columns (products). These metrics provide, at least in principle, a quantitative assessment of the capability content of an economy (that is, the economic complexity of a territory) and of a product (that is, the complexity of a product or industry).
As mentioned above, comparing territorial complexity estimates is complicated by several issues; we therefore begin by examining product complexity metrics. In this case, the products or industries under comparison are highly similar, and identical when the same classification system is used. The outputs of the algorithms should, in principle, coincide, as they are all intended to measure the capability content of a given product. As we will show, however, these estimates are mutually inconsistent.
Before analyzing these measures, Table \ref{tab:methods_desc} reports the set of product complexity indicators included in our comparison.

\begin{table}[t]
\caption{Different measures of the economic complexity of products and industries.}
\label{tab:methods_desc}
\centering
\renewcommand{\arraystretch}{1.2}
\setlength{\tabcolsep}{4pt}
\begin{tabular}{p{0.23\linewidth}p{0.28\linewidth}p{0.42\linewidth}}
\toprule
\rowcolor{orange!30}
\textbf{Name} & \textbf{Paper} & \textbf{Description} \\
\midrule
\rowcolor{orange!15}
Complexity
& Tacchella et al. (2012)
& Output of the Fitness-Complexity algorithm \\
\rowcolor{orange!10}
Extensive Complexity
& Tacchella et al. (2012)
& Output of the extensive Fitness-Complexity algorithm \\
\rowcolor{orange!15}
PCI
& Hidalgo and Hausmann (2009)
& Output of the Method of Reflections (Eigenvector method) \\
\rowcolor{orange!10}
PRODY
&  Hausmann et al. (2007)
& Weighted average of the per-capita GDPs of the countries exporting a product \\
\rowcolor{orange!15}
Job Complexity
& Russo et al. (2025)
& Weighted average of the fitness of the jobs associated to an industry \\
\botrule
\end{tabular}
\end{table}

Table \ref{tab:data_blue} reports, for each database available to us, the product classification system and the corresponding level of territorial detail. Territories are ordered from the largest to the smallest spatial units.

\begin{table}[t]
\caption{Level of detail available in our databases. For each country, we specify the administrative units and the number of digits of products' codes.}
\label{tab:data_blue}
\centering
\renewcommand{\arraystretch}{1.2}
\setlength{\tabcolsep}{4pt}
\begin{tabular}{p{0.2\linewidth}p{0.2\linewidth}p{0.52\linewidth}}
\toprule
\rowcolor{blue!18}
\textbf{Country} & \textbf{Product Digits} & \textbf{Territories} \\
\midrule
\rowcolor{blue!10}
Brazil
& 2-4
& State-Mesoregion-Microregion-Municipality \\
\rowcolor{blue!5}
China
& 2
& Province-Prefecture-County \\
\rowcolor{blue!10}
Italy
& 2-4-6
& Region-Province \\
\botrule
\end{tabular}
\end{table}

\begin{figure}[htbp]
    \centering
    \begin{subfigure}[t]{0.48\textwidth}
        \centering
        \includegraphics[width=\linewidth]{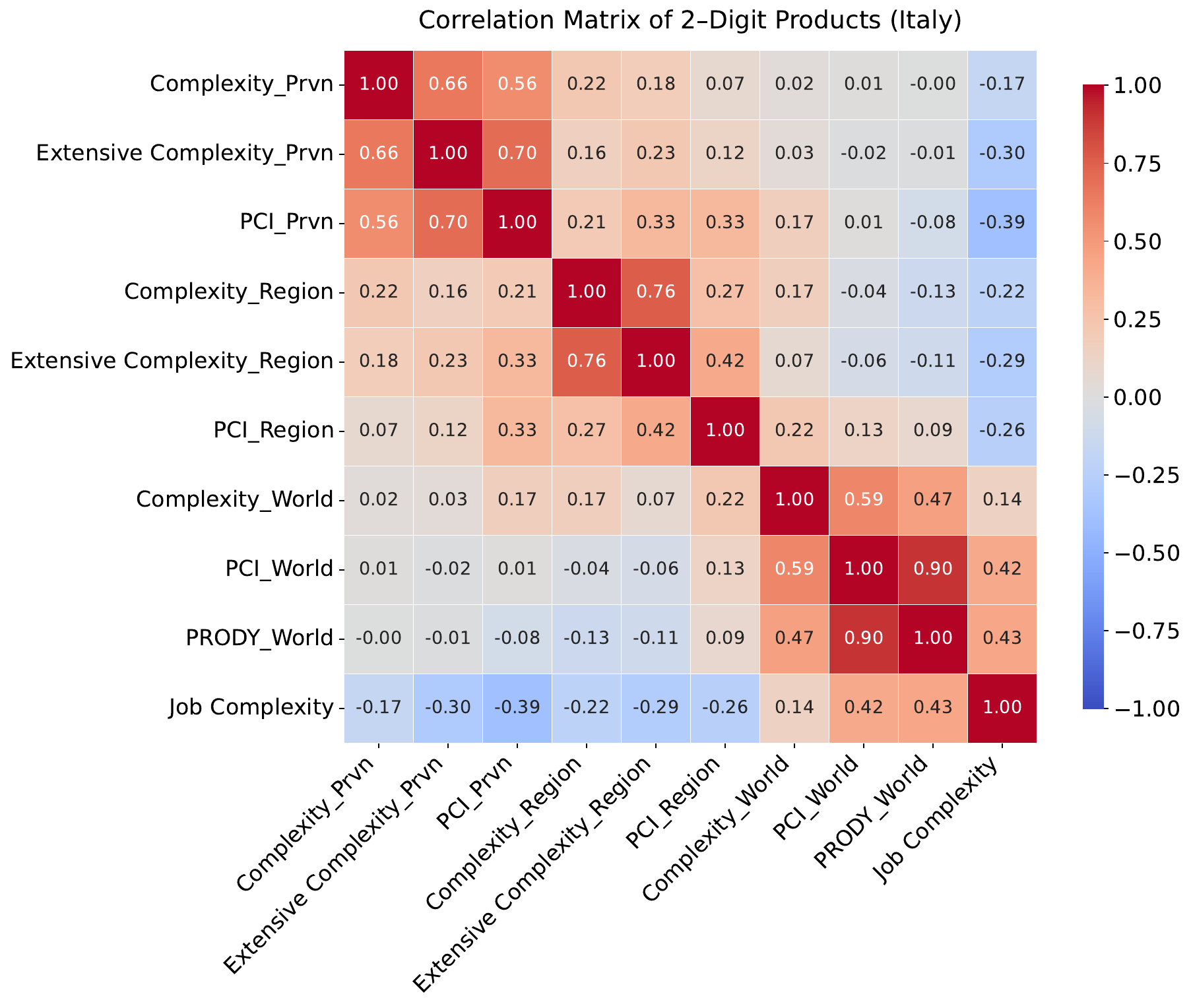}
        \caption{Italy - 2 digits}
        \label{fig:fig1b}
    \end{subfigure}
    \hfill
    \begin{subfigure}[t]{0.48\textwidth}
        \centering
        \includegraphics[width=\linewidth]{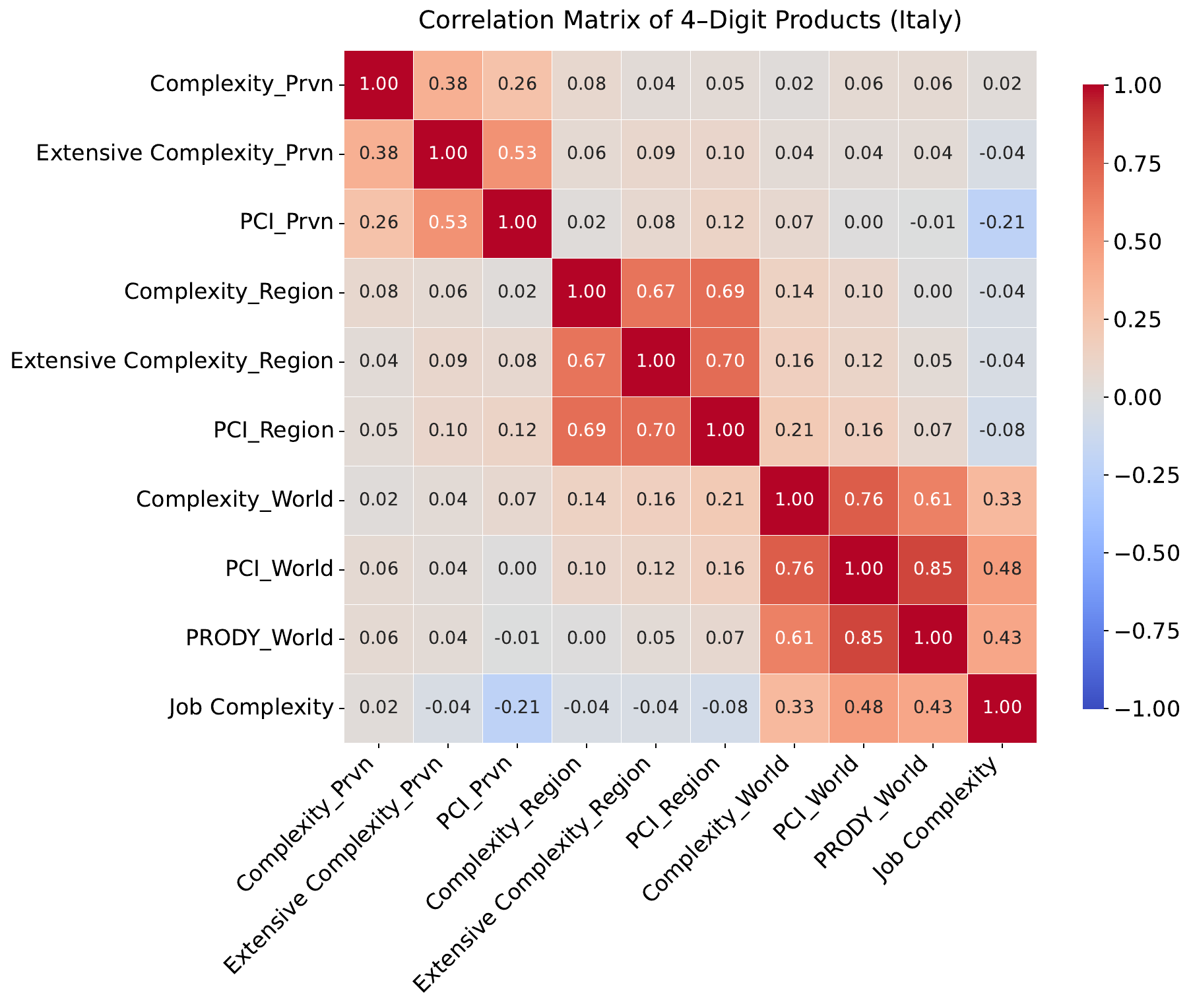}
        \caption{Italy - 4 digits}
        \label{fig:fig1c}
    \end{subfigure}

    \vspace{0.3cm}

    \begin{subfigure}[t]{0.48\textwidth}
        \centering
        \includegraphics[width=\linewidth]{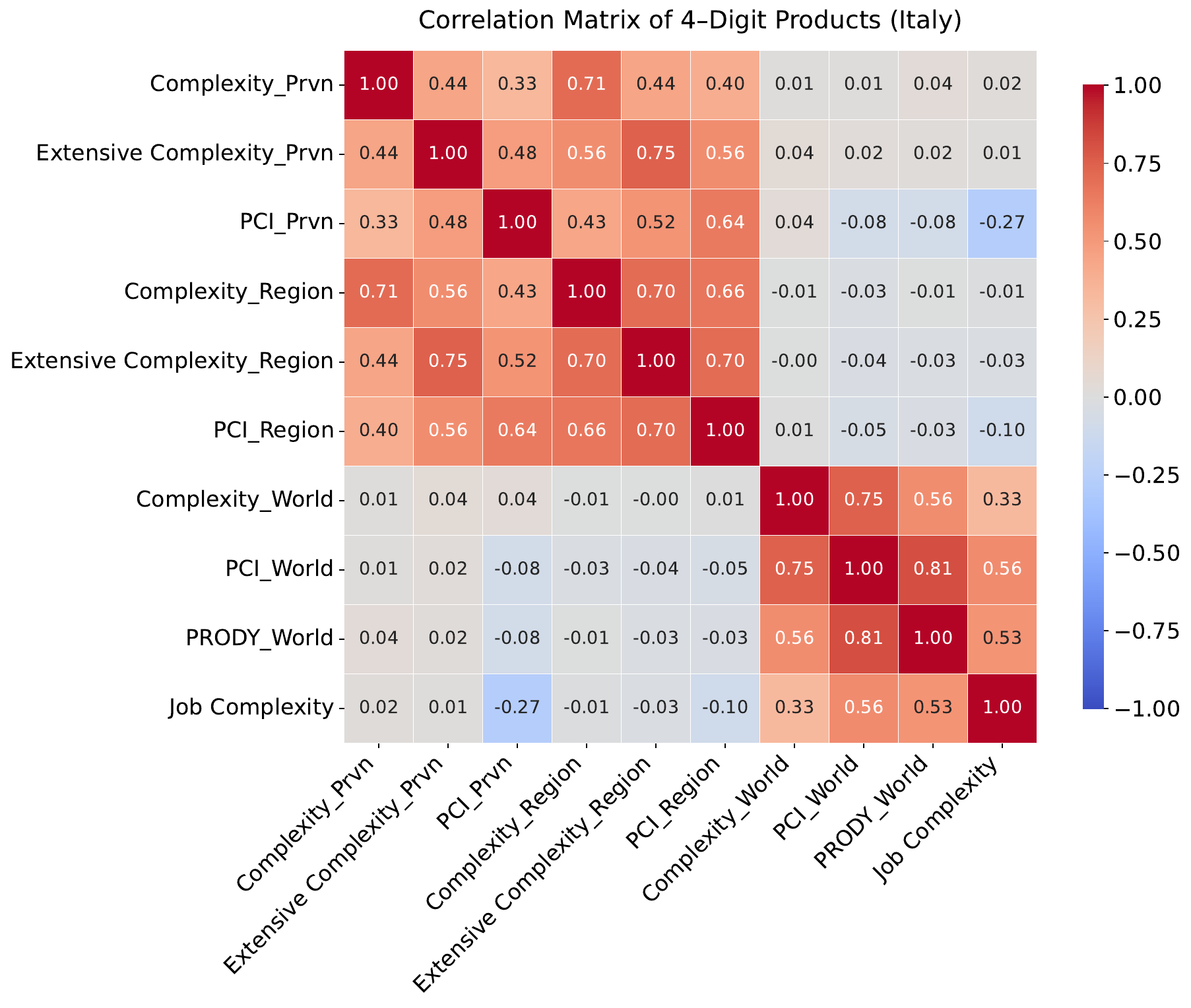}
        \caption{Italy - 6 digits}
        \label{fig:fig1d}
    \end{subfigure}
    \hfill
    \begin{subfigure}[t]{0.48\textwidth}
        \centering
        \includegraphics[width=\linewidth]{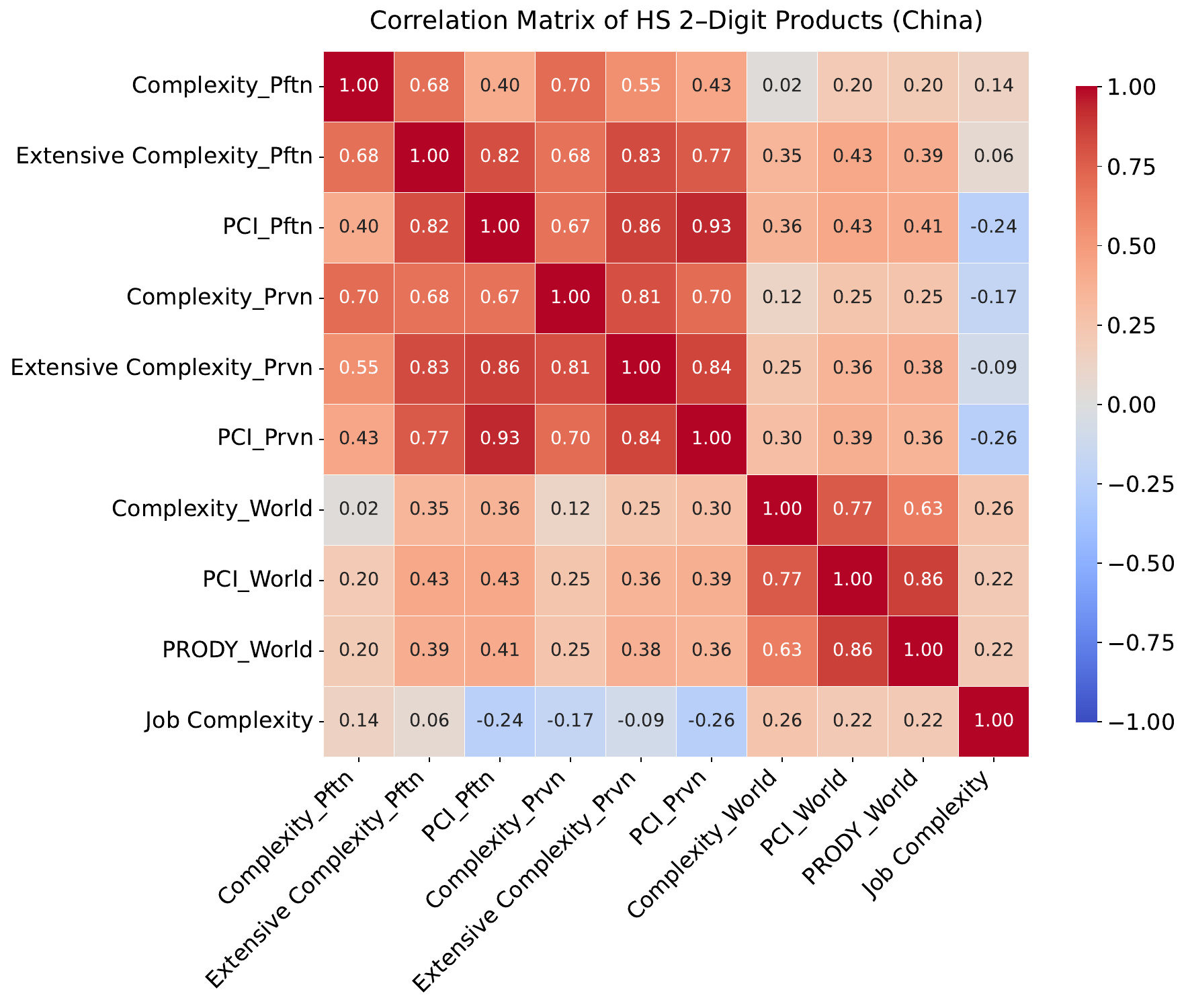}
        \caption{China - 2 digits}
        \label{fig:fig1a}
    \end{subfigure}
    \vspace{0.3cm}

    \begin{subfigure}[t]{0.48\textwidth}
        \centering
        \includegraphics[width=\linewidth]{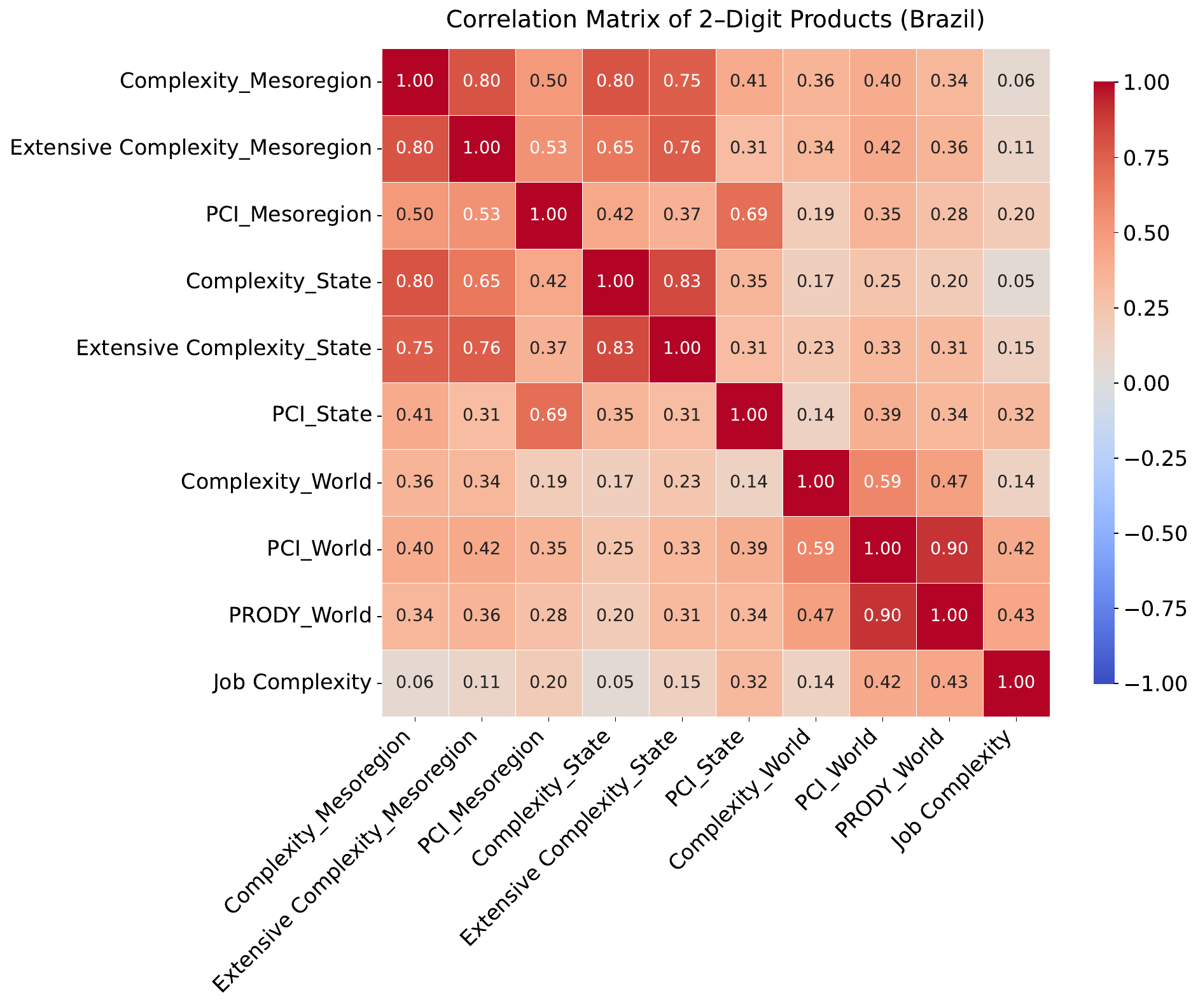}
        \caption{Brazil - 2 digits}
        \label{fig:fig1e}
    \end{subfigure}
    \hfill
    \begin{subfigure}[t]{0.48\textwidth}
        \centering
        \includegraphics[width=\linewidth]{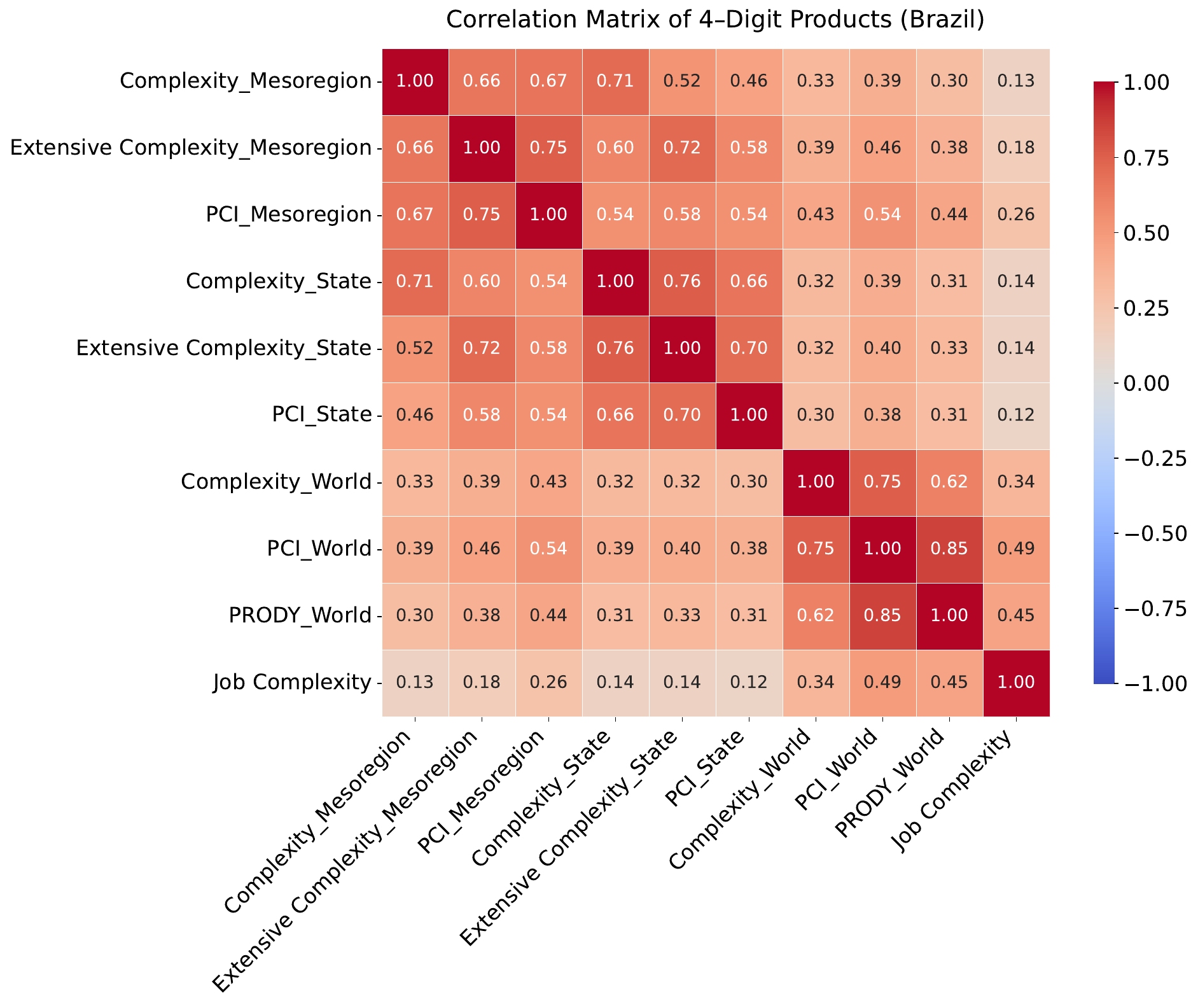}
        \caption{Brazil - 4 digits}
        \label{fig:fig1f}
    \end{subfigure}

    \caption{Correlation matrices between different metrics for the complexity of products. Each panel refers to a specific country and to a level of detail of products' classification. The correlations between the global measure are much higher than the correlations between global and local measures - thereby calling the latter into question.}
    \label{fig:product_results}
\end{figure}

In the figure, we present six Spearman correlation matrices, each referring— as indicated in the subcaptions— to a specific country and to a specific level of product classification. In all these matrices, we evaluate the same quantities: the estimates of the complexity of the same set of products, computed either from different territory–product matrices or using different algorithms, as summarized in Tables \ref{tab:methods_desc} and \ref{tab:data_blue}.

Consider, for instance, the correlation matrix in the upper-left panel, which corresponds to Italy and to products classified according to the Harmonized System at the 2-digit level (known as Chapters). The first row (and column) refers to the “Complexity” measure obtained by applying the Fitness and Complexity algorithm to the Province–Product matrix. This measure is endogenous, in the sense that the algorithm is applied directly to the subnational matrix that records the presence of 2‑digit products in each of the 109 Italian provinces. Applying different algorithms to the same territorial matrix produces the values shown in the second and third rows of the correlation matrix.

Rows three through six refer to the same algorithms, but applied endogenously to the Region–Product matrix, where the territorial units are the 20 Italian regions. Rows seven and eight instead report the complexity values computed from the country–product matrix. Row nine corresponds to PRODY, the productivity index introduced by Hausmann et al. (2007)~\cite{hidalgo2007product}, defined as the weighted average of the GDP per capita of the countries that export a given product. The final row reports Job‑based Complexity, computed from the complexity of the occupations associated with each product, as described in Russo et al. (2025)~\cite{russo2025jobbased}.

We first evaluate the general structure of the correlation matrices shown in Figure 1. A clear block structure emerges: measures are strongly correlated when computed using the same territorial level, but these correlations vanish when comparing different territorial scales or local estimates with country‑level assessments. This finding confirms and extends the result of Laudati~\cite{laudati2023ecosystems}, who documented an absence of correlation between the outcomes of economic complexity algorithms when applied at the firm level versus the country level.

This loss of correlation stands in sharp contrast with the theoretical construction underlying economic complexity algorithms: if high complexity were indeed tracing the presence of territorial capabilities, such a relationship should hold independently of the geographical scale. Instead, global measures are not correlated with local ones, revealing a substantial inconsistency both theoretically (capability‑based interpretations should not depend on the spatial scale) and empirically (we can no longer determine, in practice, which products should be considered complex).

\begin{figure}[htbp]
    \centering

    \begin{subfigure}[b]{0.48\textwidth}
        \centering
        \includegraphics[width=\linewidth]{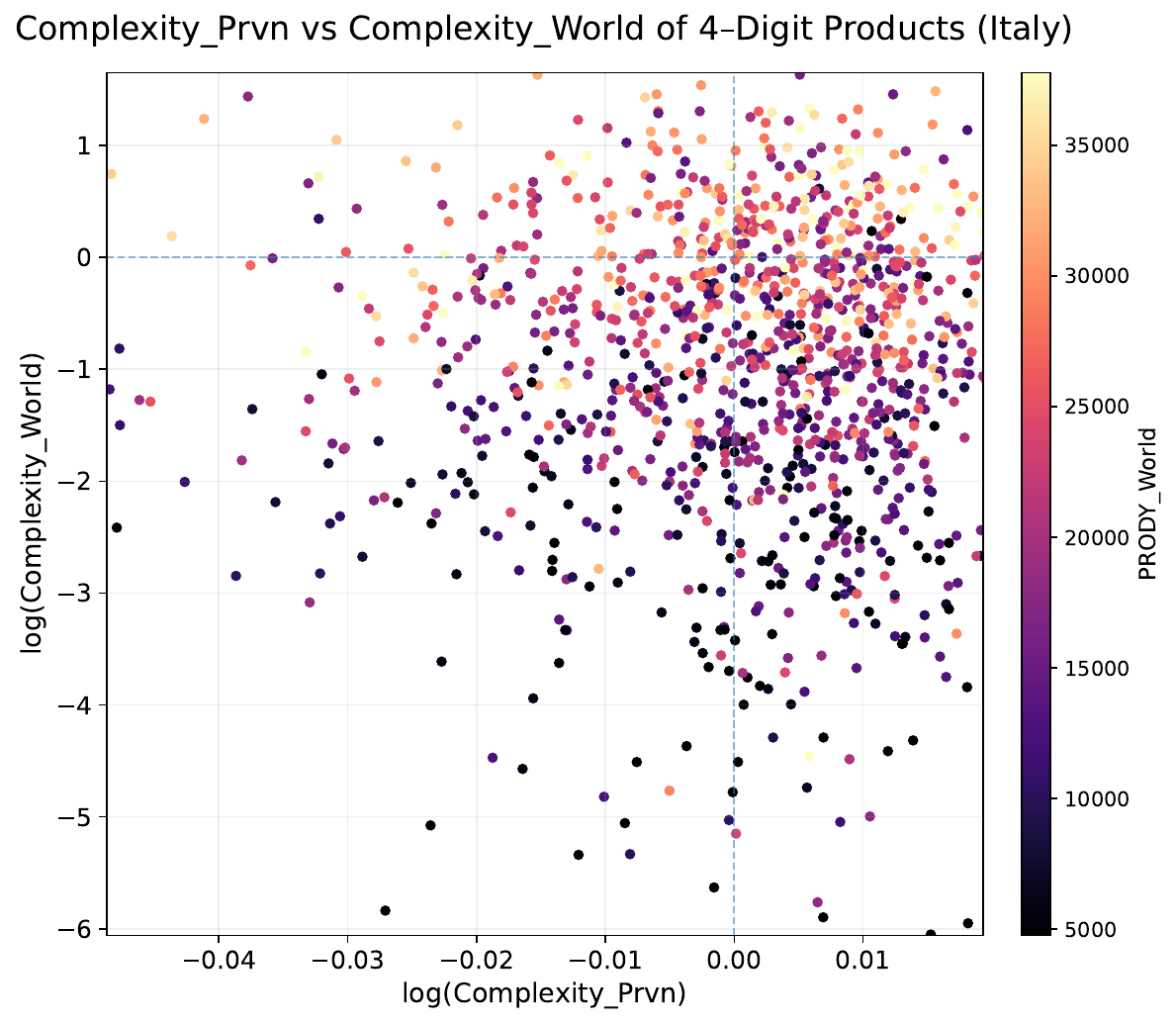}
        \label{fig:fig0a}
    \end{subfigure}
    \hfill
    \begin{subfigure}[b]{0.48\textwidth}
        \centering
        \includegraphics[width=\linewidth]{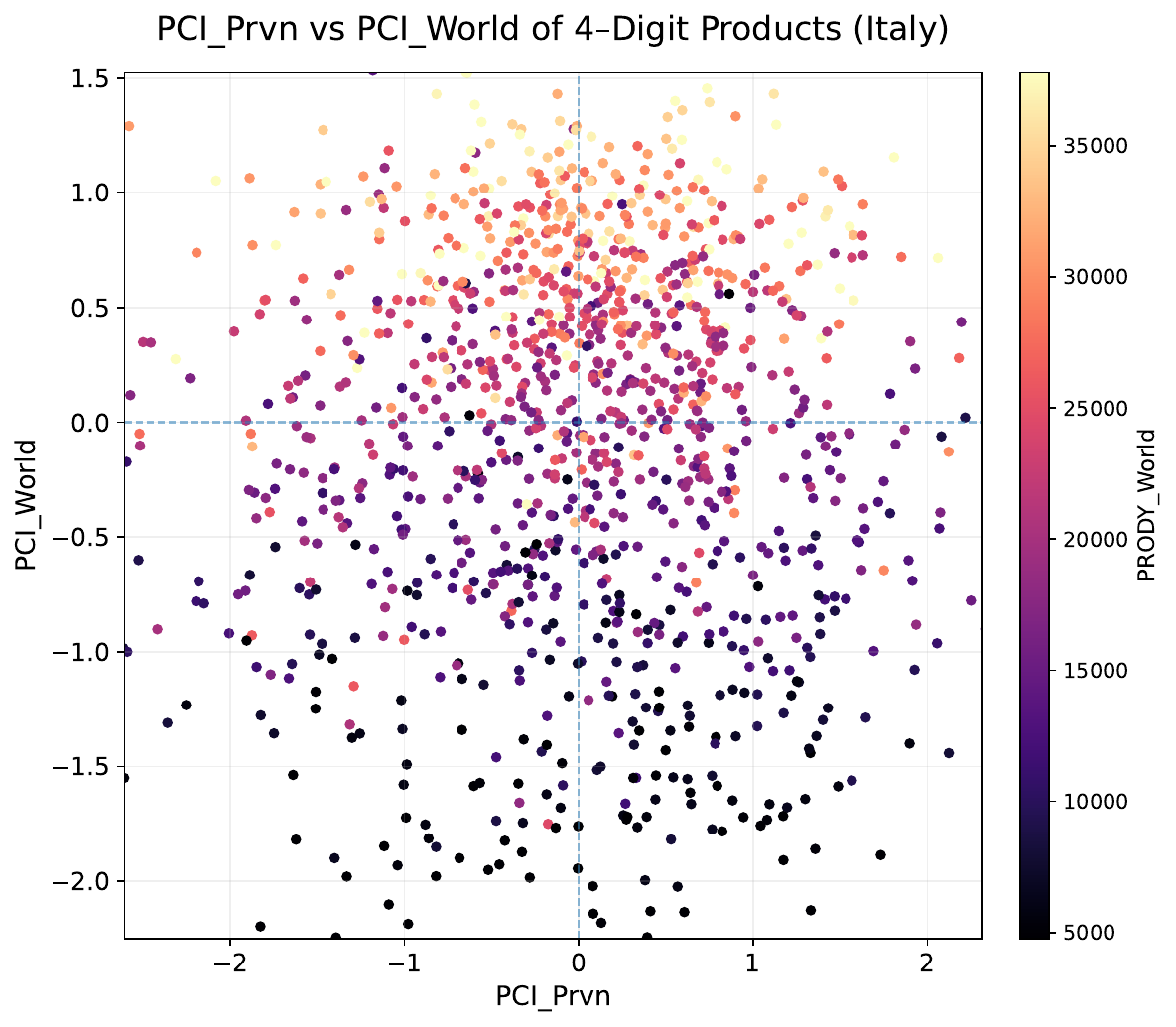}
        \label{fig:fig0b}
    \end{subfigure}

    \caption {Comparison between the same product complexity measures (Complexity on the left, PCI on the right) computed using international trade data (y‑axis) and subnational data (x‑axis). The absence of correlation is evident. Point colors represent the PRODY index, a simple measure of industrial sophistication based on the GDP per capita of exporting countries. PRODY is clearly correlated with global complexity measures but not with local ones, thereby casting doubt on the latter.}
    \label{fig:scattercomp}
\end{figure}

Our conclusion is that economic complexity algorithms cannot—and should not—be applied at the subnational level. On the contrary, the globally computed measures are not only mutually correlated but are also aligned with independent indicators such as PRODY and Job‑based Complexity, which substantiates the robustness of the global assessment. This is evident from Figure \ref{fig:scattercomp}, in which we compare the same two measures (Complexity from the EFC method and PCI) computed using world data (y axis) and subnational data (x axis). The absence of correlation is evident. The color is given by the PRODY value, whose gradient clearly develops along the vertical direction.

\clearpage
\subsection{Subnational economic complexity: which metric to use?}
\begin{table}[t]
\caption{Different measures of the economic complexity of territories.}
\label{tab:terr_metrics}
\centering
\renewcommand{\arraystretch}{1.2}
\setlength{\tabcolsep}{4pt}
\begin{tabular}{p{0.30\linewidth}p{0.65\linewidth}p{0.42\linewidth}}
\toprule
\rowcolor{orange!30}
\textbf{Name} & \textbf{Description} \\
\midrule
\rowcolor{orange!15}
Fitness
& Output of the Fitness-Complexity algorithm (binary matrix)\\
\rowcolor{orange!10}
Exogenous Fitness (sum)
& Sum of World Complexities \\
\rowcolor{orange!15}
Exogenous ECI (sum)
& Sum of World PCI \\
\rowcolor{orange!10}
Extensive Fitness
&  Output of the Fitness-Complexity algorithm (weighted matrix)\\
\rowcolor{orange!15}
Exogenous Extensive Fitness
& Weighted sum of World Complexities \\
\rowcolor{orange!10}
Job-based Extensive Fitness
& Weighted sum of Job-Based Complexities \\
\rowcolor{orange!15}
ECI
& Output of the Method of Reflections (Eigenvector method) \\
\rowcolor{orange!10}
Exogenous Fitness (average)
& Average of World Complexities \\
\rowcolor{orange!15}
Exogenous ECI (average)
& Average of World PCIs \\
\botrule
\end{tabular}
\end{table}

\begin{figure}[htbp]
    \centering

    \begin{subfigure}[t]{0.48\textwidth}
        \centering
        \includegraphics[width=\linewidth]
        {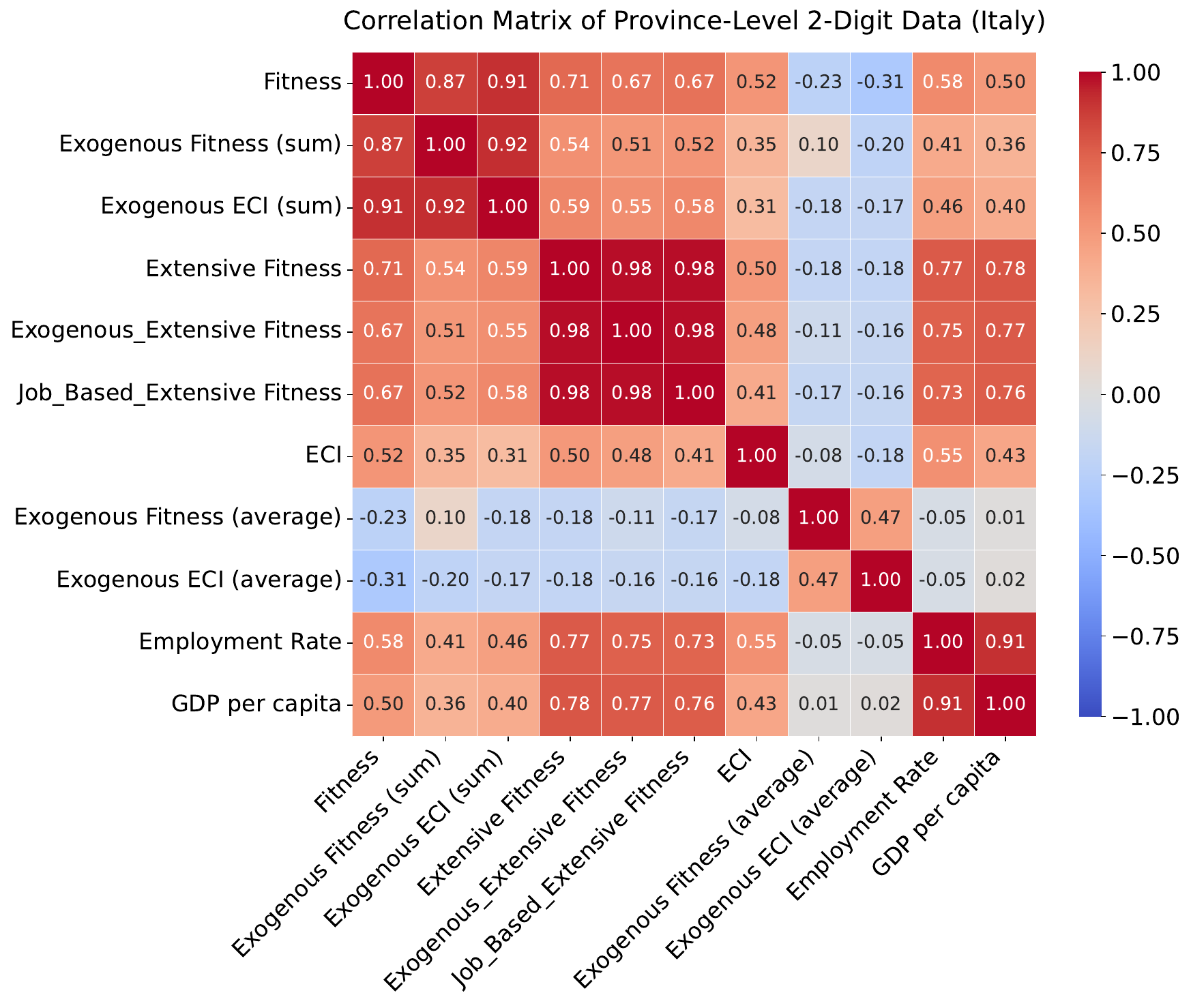}
        \label{fig:fig2a}
    \end{subfigure}
    \hfill
    \begin{subfigure}[t]{0.48\textwidth}
        \centering
        \includegraphics[width=\linewidth]
        {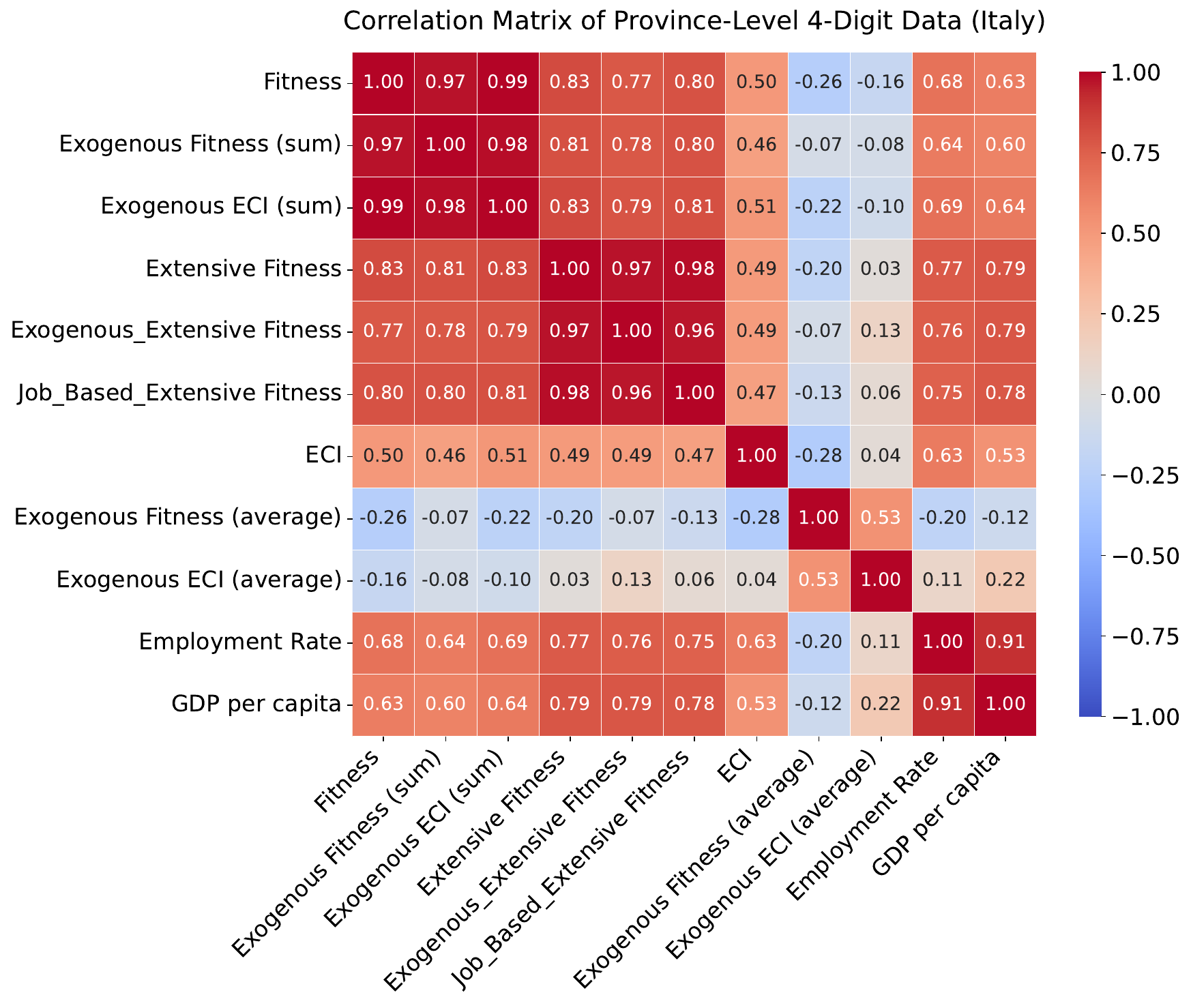}
        \label{fig:fig2b}
    \end{subfigure}

    \vspace{0.3cm}

    \begin{subfigure}[t]{0.48\textwidth}
        \centering
        \includegraphics[width=\linewidth]
        {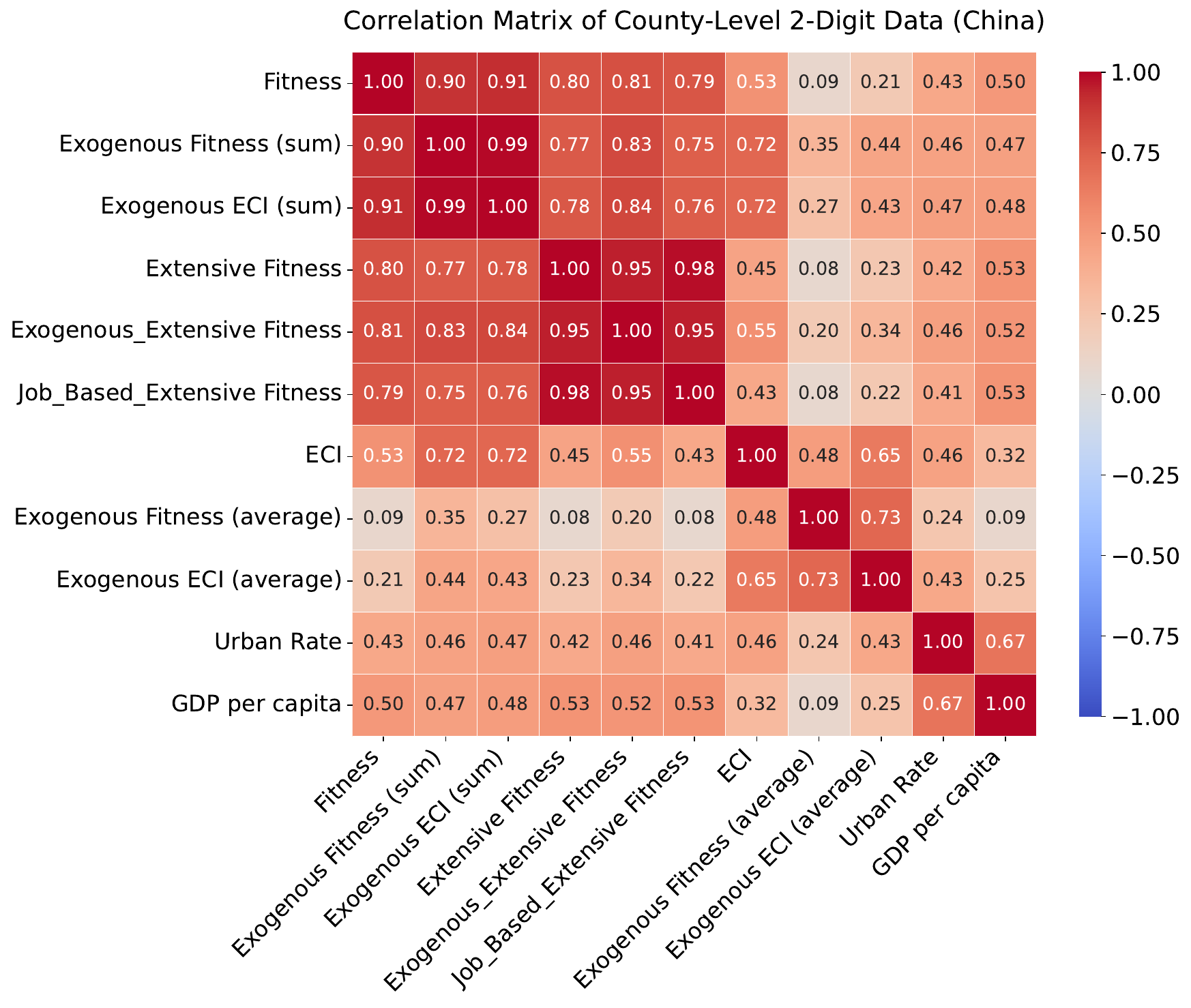}
        \label{fig:fig2c}
    \end{subfigure}
    \hfill
    \begin{subfigure}[t]{0.48\textwidth}
        \centering
        \includegraphics[width=\linewidth]{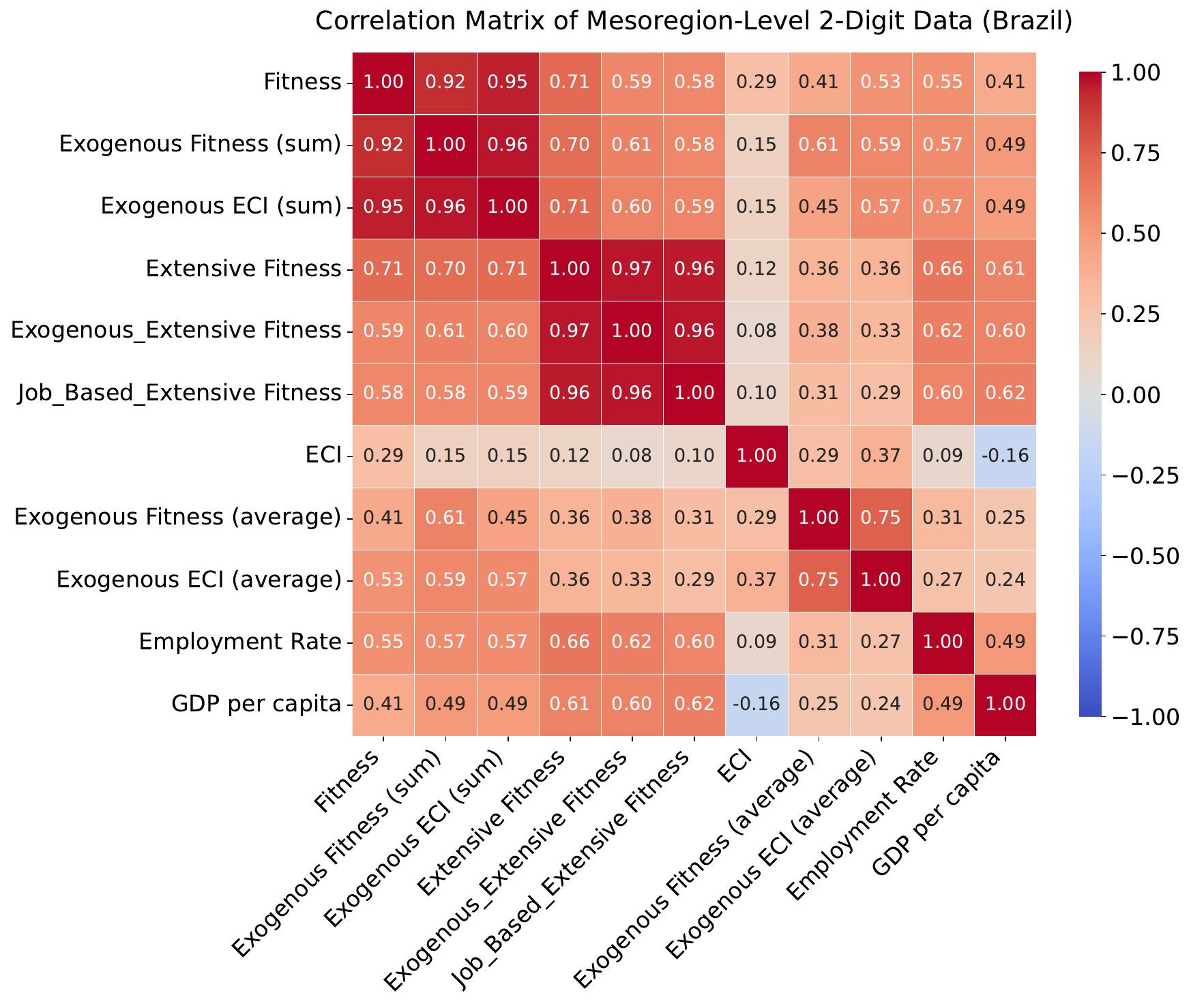}
        \label{figeeee}
    \end{subfigure}

    \caption{Correlation matrices between different metrics for the economic complexity of territories. Each panel refers to a specific country, to a geographical scale, and to a level of detail of products’ classification.
    We find a high correlation between similar measures (sum or average-based). Exogenous + extensive measures show a higher correlation with economic indicators.}
    \label{fig:admin_results_4}
    
\end{figure}

We now turn to the analysis of territorial metrics, that is, to comparing estimates—obtained through different techniques—of the same administrative units. As in the previous section, we present the results using correlation matrices, shown in Figure \ref{fig:admin_results_4}. Table \ref{tab:terr_metrics} provides a brief description of the metrics considered, whose computation is detailed in the Methods section.

The first three metrics are constructed as sums over products derived from the binarized matrix. Consequently, they reward the most diversified territories. As shown, these metrics are all highly correlated with each other, regardless of the specific weight (i.e., product‑level complexity measure) assigned to each product.

Moving down the rows of the matrices, we encounter three extensive metrics, constructed instead from the weighted matrices. In this case, the territory–product matrix takes continuous values between 0 and 1, corresponding to the fraction of exported value, analogous to the market share used at the country level \cite{tacchella2012fitness,patelli2022MS}. These three metrics are likewise highly correlated with one another.

The seventh row reports ECI, the measure obtained by applying the algorithm introduced by Hidalgo and Hausmann \cite{hidalgo2009building} to the territory–product matrix. Note that this estimate of economic complexity is endogenous, like Fitness (row 1) and Extensive Fitness (row 4), whereas the other metrics are exogenous, meaning that they rely on complexity values computed globally rather than locally. Interestingly, ECI is fairly well correlated with all the measures: those based on diversification, those based on diversification weighted by market share, and those based on averaging, which appear in rows 8 and 9 of the correlation matrices.

These latter two metrics are computed by averaging, for each territory, the globally computed values of Complexity and PCI. What emerges clearly is that these metrics capture different aspects of territorial structure. Strikingly, in some cases they are even negatively correlated, as observed for the averaging‑based metrics in the case of Italy.

Consequently, in the absence of self‑consistency criteria—such as those available for product complexity measures—one is left with the question of which metric best captures the concept of economic complexity. To address this, we compare, within the same correlation matrices, the economic complexity metrics (rows 1–9) with two standard economic indicators: GDP per capita and the employment rate. We consider that a positive association should exist between the presence of capabilities in a territory, on the one hand, and the production of goods and services and the corresponding number of workers, on the other.
It is important to note that our objective is not to establish a statistically significant relationship between these variables, nor to infer any causal link. This is typically the aim of studies in the existing literature, where linear regressions and appropriate control variables are employed. In those studies, a single economic complexity metric, a single geographical scale, a single country, and a single product classification are used—implicitly assuming that these choices do not affect the results.
In contrast, our goal here is to understand the implications of these methodological choices by directly comparing the results obtained when different choices are made.

Inspection of Figure \ref{fig:admin_results_4} shows that, on average, the unweighted diversification‑based measures (rows 1–3) are moderately correlated with the economic indicators, as is ECI (with the exception of Brazil). By contrast, the averaging‑based indicators (rows 8 and 9) display weak correlations. The metrics that exhibit the strongest associations with both GDP per capita and the employment rate are those based on weighted sums—that is, the extensive indicators. Paradoxically, these are the least commonly used measures in the literature, where binarization of the matrix via the Revealed Comparative Advantage is typically preferred, despite the fact that, even from a theoretical standpoint, its use is more appropriate at the country level (given that competition among subnational export baskets is not analogous to competition among national economies).
For reasons of space, we do not include in the main text the correlation matrices corresponding to other administrative units or to the additional levels of product detail available. Interested readers may find them in the Supplementary Material.

\begin{figure}[htbp]
    \centering

    \begin{subfigure}[t]{0.48\textwidth}
        \centering
        \includegraphics[width=\linewidth]
        {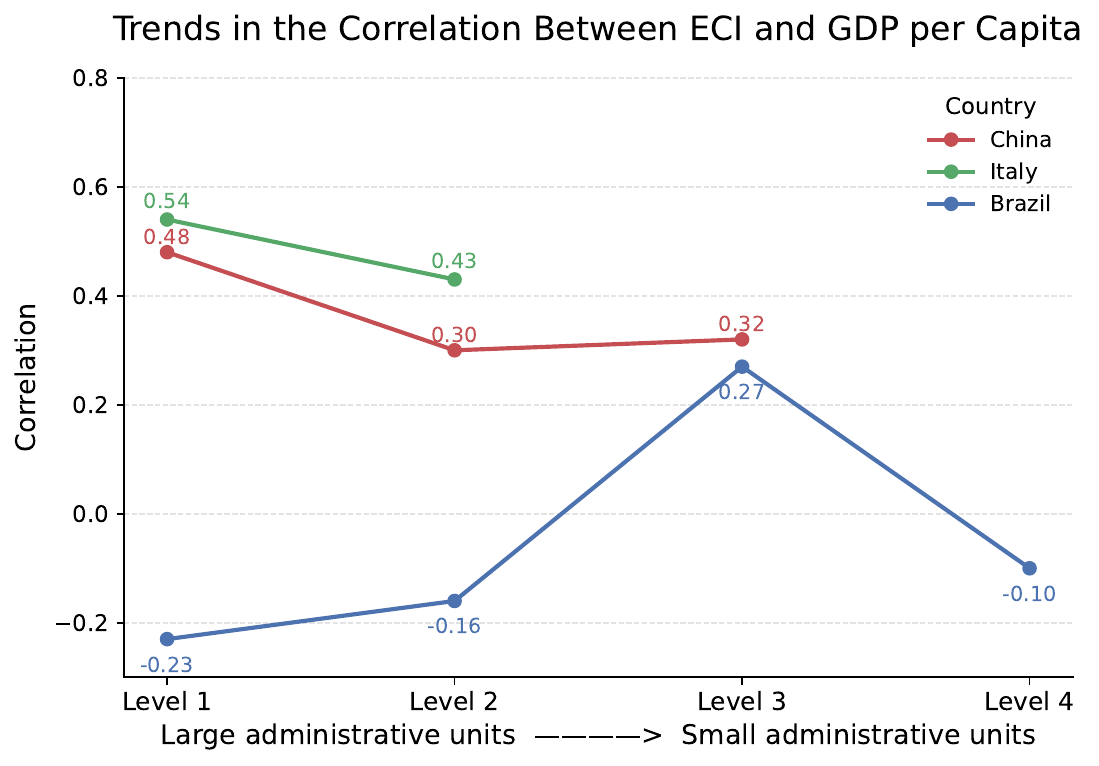}
        \caption{Spearman correlation between ECI and GDP per capita}
        \label{fig:fig3a}
    \end{subfigure}
    \hfill
    \begin{subfigure}[t]{0.48\textwidth}
        \centering
        \includegraphics[width=\linewidth]
        {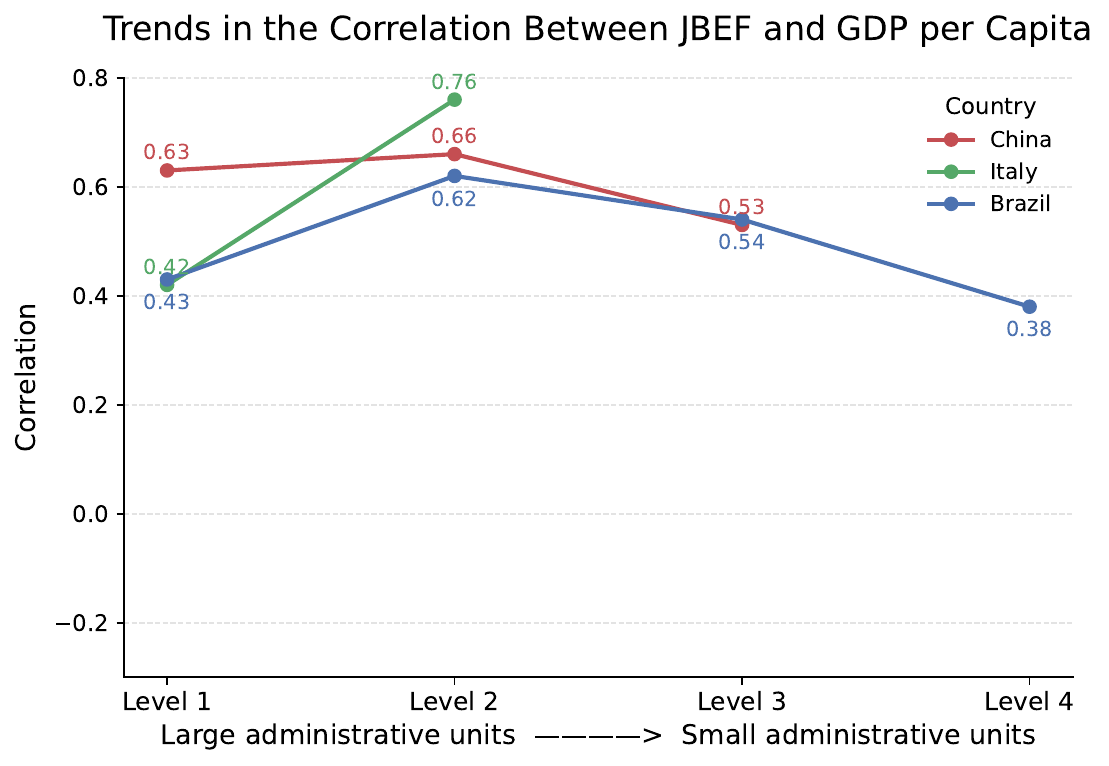}
        \caption{Spearman correlation between Job Based Extensive Fitness and GDP per capita}
        \label{fig:fig3b}
    \end{subfigure}

    \caption{Correlation trends between Economic complexity measurement indicators and GDP per capita across administrative levels and countries at 2-digit level. 
    For China, Italy, and Brazil, economic output is measured by GDP per capita.}
    
    \label{fig:figure3}

\end{figure}

In Figure \ref{fig:figure3}, we visualize how the correlation between GDP per capita and two economic complexity metrics (the Economic Complexity Index and Job‑Based Extensive Fitness) varies as a function of geographical scale. This allows us to assess, for the three countries considered, the consistency of the measures when moving from larger to smaller administrative units. The left panel reports the correlation between ECI and GDP per capita. In highly industrially diversified countries such as China and Italy, the correlation at the highest administrative level is strongly significant. This is consistent with findings in the literature, and in particular with the results of Mealy et al. \cite{mealy2019interpreting}.
In contrast, for Brazil the correlation is negative, which stands in sharp contradiction with the notion that ECI should in some way capture the capability content of a territory. To explain this result, we recall—following \cite{mealy2019interpreting}—that ECI is effectively equivalent to a clustering algorithm that partitions territories and products into two internally coherent groups. Naturally, the identification of which cluster is the “upper” one and which is the “lower” one is entirely arbitrary; consequently, ECI values are determined only up to sign.
In the official library
\footnote{https://github.com/harvard-growth-lab/py-ecomplexity/blob/master/ecomplexity/ecomplexity.py}
the sign of ECI is chosen such that it correlates positively with diversification; we follow the same convention. The issue is that this choice results in a positive correlation with GDP per capita only if administrative units span a sufficiently wide range of diversification levels—as is the case for Italy and China. For Brazil, however, differences in diversification across geographic units are relatively small, with a highly concentrated distribution, and the most diversified regions do not necessarily correspond to the most economically developed ones. 


The right panel shows the correlation between GDP per capita and the measure we propose in this paper, Job‑Based Extensive Fitness. This metric displays consistently high correlations across all countries and administrative levels, making it a strong candidate for representing economic complexity across different geographical scales.

\clearpage
\section{Methods}\label{sec11}
This section describes the data and methodologies employed to generate the results presented in the previous section.
\subsection{Databases}
The comparison of local economic complexity metrics is based on export and production data collected at a high level of geographical detail. In this study, we use data from three countries that differ markedly in terms of their level of development and industrial organization, with the aim of producing results that are as generalizable as possible. In all three cases, the output of the data preprocessing procedure is a matrix \(X\)  that quantifies industrial activity in a given economic sector for a given territory.\\
Additionally, in this paper we use data about GDP per capita from GDP and population data, the employment rate from employment counts and total population, and the urbanization rate from the urban resident population and total population. For brevity, these constructions are not detailed further here.
\subsubsection{Brazil}
The data on territorial economic activity for Brazil are obtained from the Dataviva platform (\url{https://legacy.dataviva.info/en/data/}), which provides HS 4-digit export data across different administrative levels. We use the average data from 2012 to 2022, excluding 2021 due to data unavailability. We retain only those products that are exported by at least two administrative units.

The final dataset for Brazil includes 27 state-level units, 134 mesoregions, 476 microregions, and 1,663 municipalities. We construct the \(X\) matrix for Brazil, where each row represents an administrative unit and each column corresponds to an HS product. The matrix entries represent the total export value of each product.

Employment data are also obtained from Dataviva. GDP data are computed as the average over the period 2012–2022, while population data are based on the average of 2010 and 2022 values, both sourced from the Brazilian Institute of Geography and Statistics (IBGE) (\url{https://www.ibge.gov.br/}).

The HS 4-digit classification includes approximately 1,180 products, which can be aggregated into 97 products under the HS 2-digit classification.

\subsubsection{China}
This classification also includes services, as the original data are sourced from the China Stock Market \& Accounting Research Database (CSMAR) (\url{https://data.csmar.com/}), which provides information on registered companies from 2018 to 2022. We use firms' registered capital to construct the \(X\) matrix for China.

In this matrix, each row represents a firm, and each column corresponds to a two-digit industry code based on GB/T 4754-2017. The matrix entries are given by firms' registered capital. Observations with registered capital below 1,000 are excluded. The final dataset for China includes more than 2,500 counties, approximately 330 cities, and 31 provincial-level units.

To link global production data with China's industry classification (GB/T 4754-2017), we rely on its official correspondence with the International Standard Industrial Classification of All Economic Activities (ISIC Rev.4). We then use the HS–ISIC correspondence table provided by the Jordan Industrial Observatory (\url{https://observatory.huzaifazoom.org/blog/2022/04/18/hs-isic-correspondence-table/}) to map HS-based production data to the GB/T 4754-2017 classification. Since the HS classification does not include services, a total of 53 two-digit GB/T 4754-2017 industry categories are matched.

For macroeconomic variables, the data are primarily obtained from the China National Research Data Service Platform (CNRDS)(\url{https://www.cnrds.com/Home/Login}). For some key missing data, such as Beijing and Shanghai, we manually supplement the dataset using the corresponding Statistical Yearbooks.

Population data are derived from the China Population Census published by the National Bureau of Statistics of China. Specifically, we use the average population between 2010 and 2020 to compute per capita measures.

\subsubsection{Italy}
The data are sourced from the Italian National Institute of Statistics (ISTAT) and are documented at (\url{https://esploradati.istat.it/coeweb/databrowser/}). The dataset contains information on export volumes for goods shipped abroad by nearly the entire population of Italian firms in 2017, totaling over one million distinct exporters.  For each observation, the data identify the exporting firm, the product exported, the year of export, and the corresponding value. Products are classified according to the six-digit Harmonized System (HS) classification, yielding a total of 5,203 unique product categories. This fine-grained product classification allows for a detailed reconstruction of firms’ export baskets and supports the aggregation of exports across different geographical levels. Specifically, Italy is subdivided administratively into
20 regions units and 109 provinces, which are used in our spatial analysis.
\subsection{Data preprocessing}\label{sec:dataprep}
Based on the administrative unit--product export matrix $\textbf{X}$, we further construct two types of matrices using the revealed comparative advantage (RCA) method \cite{balassa1965trade} and the market share (MS) method \cite{tacchella2012fitness,patelli2022MS}, respectively. The RCA method generates a binary matrix, whereas the MS method generates a weighted matrix. Let $X_{cp}$ denote the export value of product $p$ by administrative unit $c$. The RCA of unit $c$ in product $p$ is defined as

\begin{equation}
RCA_{cp} =
\frac{X_{cp}/\sum_p X_{cp}}
{\sum_c X_{cp}/\sum_c \sum_p X_{cp}}.
\label{eq:rca}
\end{equation}

Here, $\sum_p X_{cp}$ is the total export value of all products by unit $c$, $\sum_c X_{cp}$ is the total export value of product $p$ across all administrative units, and $\sum_c \sum_p X_{cp}$ is the total export value of all products across all administrative units. The binary matrix element $M_{cp}$ is then defined as

\begin{equation}
M_{cp} =
\begin{cases}
1, & \text{if } RCA_{cp} > 1, \\
0, & \text{otherwise}.
\end{cases}
\label{eq:binary_matrix}
\end{equation}

In addition, the market share method is used to construct a weighted matrix. The market share of unit $c$ in product $p$ is defined as
\begin{equation}
MS_{cp} = \frac{X_{cp}}{\sum_c X_{cp}}.
\label{eq:ms}
\end{equation}

Unlike the RCA method, which produces a binary matrix, the MS method directly generates a weighted matrix, whose entries vary between 0 and 1.

\subsection{Indicators of economic complexity}
In this subsection we describe how, starting from the matrices \textbf{M} and \textbf{MS} described above, we compute the various assessments of the economic complexity of both products and territories. 

\subsubsection{PRODY}\label{subsec7}
One of the first assessments of the sophistication of a product was introduced by Hausmann et al.~\cite{hausmann2007export}. The PRODY of product $p$ is defined as a weighted average of countries’ per-capita GDP, where the weight $w_{cp}$ for country $c$ is the share of product $p$ in country $c$’s total exports $X_c = \sum_p X_{cp}$, normalized across all countries. Denoting per-capita GDP by ${Y}_c$, the construction is given in Eq.~(\ref{eq:Pr}).

\begin{equation}
\left\{
\begin{aligned}
w_{cp} &= 
\frac{X_{cp}/X_c}
{\sum_{c'} X_{c'p}/X_{c'}},
\qquad \sum_c w_{cp}=1, \\
PRODY_p &= \sum_c w_{cp} Y_c .
\end{aligned}
\right.
\label{eq:Pr}
\end{equation}

Note that the weight normalization is conducted product by product, so that the weights sum to one across countries for each product. PRODY is one of the early indicators that highlighted the important relationship between a country’s export structure and its level of economic development. The idea is that, on average, high productivity levels/sophisticated products are associated with high-income countries. Its simple formulation allows its use as a benchmark for complexity measures, as done in \cite{laudati2023ecosystems}.

\subsubsection{Economic Complexity Index}\label{subsec4}
The Economic Complexity Index (ECI) \cite{hidalgo2009building} is algorithmically constructed from the territory--product bipartite network by refining, at each iterative step, the diversity (or diversification) of territories and the ubiquity of products. In its original formulation, \textbf{M} denotes the starting binary matrix, where $M_{cp}=1$ if country $c$ has a revealed comparative advantage in product $p$ (that is, if $RCA_{cp}>1$), and $M_{cp}=0$ otherwise. The diversity of a country and the ubiquity of a product are respectively defined as
\begin{equation}
k_{c,0} = \sum_p M_{cp},
\qquad
k_{p,0} = \sum_c M_{cp}.
\end{equation}
Following the Method of Reflections \cite{hidalgo2009building}, the higher-order quantities are iteratively defined as
\begin{equation}
k_{c,n} = \frac{1}{k_{c,0}} \sum_p M_{cp} k_{p,n-1},
\qquad
k_{p,n} = \frac{1}{k_{p,0}} \sum_c M_{cp} k_{c,n-1}.
\end{equation}
This iterative process can be reformulated as an eigenvalue problem \cite{cristelli2013measuring}. The country--country matrix is defined as
\begin{equation}
\tilde{M}_{cc'} = \frac{1}{k_{c,0}} \sum_p \frac{M_{cp} M_{c'p}}{k_{p,0}},
\end{equation}
from which one can define the eigenvector $K_c$ satisfying
\begin{equation}
\sum_{c'} \tilde{M}_{cc'} K_{c'} = \lambda K_c.
\end{equation}
ECI corresponds to the eigenvector associated with the second largest eigenvalue (the one associated with the first being trivial). Analogously, one can define the Product Complexity Index (PCI) as the eigenvector associated to the second largest eigenvalue of the product--product matrix. To ensure comparability across regions, the ECI values are standardized using z-score normalization. Importantly, as our analysis is based on Spearman rank correlations, which depend only on rank information, the results are unaffected by this transformation.
Note that both ECI and PCI are standardized as
\begin{equation}
ECI_c = \frac{K_c - \langle K \rangle}{\mathrm{std}(K)}.
\end{equation}

Finally, note that this approach can be interpreted as a clustering problem \cite{mealy2019interpreting}; in particular, the sign of the eigenvectors labels the two clusters, and positive ECI countries are associated with positive PCI products. In this situation, the sign of ECI can not be unambiguously defined. Coherently with the original Method of Reflections, we fixed the sign of ECI in such a way that its correlation with the diversification is positive.

\subsubsection{Economic Fitness and Complexity}\label{subsec1}
The Economic Fitness and Complexity (EFC) \cite{tacchella2012fitness} algorithm is defined through a nonlinear iterative system, as given in Eqs.~(\ref{eq:efc1})--(\ref{eq:efc2}). By iterating these equations until convergence, one obtains the fitness $F_c$ of country $c$ and the complexity $Q_p$ of product $p$. As before, $M_{cp}$ denotes the binary country (or territory)--product matrix, where $M_{cp}=1$ if country $c$ is competitive in product $p$, and $M_{cp}=0$ otherwise. At each iteration, both fitness and complexity are normalized by their corresponding averages to ensure numerical stability.
\begin{align}
\left\{
s\begin{aligned}
F_c^{\prime (n)} &= \sum_p M_{cp} \, Q_p^{(n-1)}, \\
Q_p^{\prime (n)} &= \frac{1}{\sum_c M_{cp} \, \frac{1}{F_c^{(n-1)}}}
\end{aligned}
\right.
\label{eq:efc1}
\\[8pt]
\left\{
\begin{aligned}
F_c^{(n)} &= \frac{F_c^{\prime (n)}}{\langle F^{\prime (n)} \rangle_c},\\
Q_p^{(n)} &= \frac{Q_p^{\prime (n)}}{\langle Q^{\prime (n)} \rangle_p}
\end{aligned}
\right.
\label{eq:efc2}
\end{align}
The initial conditions are set as $F_c^{\prime (0)} = 1$ for all countries $c$ and $Q_p^{\prime (0)} = 1$ for all products $p$. The binary matrix element $M_{cp}$ is defined in Section \ref{sec:dataprep}. In this paper, we used the above approach to compute the complexity $Q_p$ of products using world trade data. These values are then used to compute the exogenous Fitness and ECI at the territorial level, as described in Section \ref{sec:exo}. When implementing the above methodology directly at the subnational level, we have found the convergence issues described in \cite{pugliese2016convergence}, so we adopted the modified version of the EFC algorithm described in the next section.

\subsubsection{The non-homogeneous EFC algorithm}\label{subsec2}

Although the original EFC method provides strong support for measuring and forecasting economic development, researchers have observed that, in certain cases, the structure of the reordered matrix \textbf{M} prevents an economy’s final fitness from reaching finite, nonzero values \cite{pugliese2016convergence}. To address this issue, Servedio et al. proposed a modified, EFC-based formulation~\cite{servedio2018fitness}. The core idea is to deform the original recursion by introducing a small regularization parameter $\delta$ to prevent Fitness from vanishing:

\begin{equation}
\left\{
\begin{aligned}
\tilde{F}_c^{(n)} &= \delta^2 + \sum_{p'} \frac{M_{c p'}}{\tilde{P}_{p'}^{(n-1)}}, 
\qquad 1 \leq c \leq C, \\
\tilde{P}_p^{(n)} &= 1 + \sum_{c'} \frac{M_{c' p}}{\tilde{F}_{c'}^{(n-1)}}, 
\qquad 1 \leq p \leq P.
\end{aligned}
\right.
\label{eq:efc_new}
\end{equation}

Here we used $\delta = 1 \times 10^{-10}$ and tested that the results are stable for lower values of $\delta$.

\subsubsection{Exogenous Fitness}\label{sec:exo}

Both the ECI and EFC methods compute their respective metrics by applying iterative algorithms to the matrix $\textbf{M}$ (which, in the original applications, is the country–product matrix). In subnational analyses, the literature typically defines $\textbf{M}$ as a territory–product matrix, where territories correspond to the local administrative units described in the Database section. In this case, the economic complexity measures are derived solely from within-country production data, and we can call this approach \textit{endogenous} economic complexity. However, as we have seen in the results section, product complexity inferred only from domestic data can diverge from complexity as computed in the global trade network (the latter assessment is endogenous, too; but $M$ is the country-product matrix). Operti et al.~\cite{operti2018dynamics} propose obtaining an \textit{exogenous} measure of economic complexity by combining product complexity values estimated from international trade with the country’s internal production-structure matrix ${M}$. The specific formulation is given in equations Eq.~(\ref{eq:Fs}).
\begin{equation}
\left\{
\begin{aligned}
\tilde{F}_s &= \sum_p M_{sp} \, Q_p, \\
F_s &= \frac{\tilde{F}_s}{\langle \tilde{F}_s \rangle_s}.
\end{aligned}
\right.
\label{eq:Fs}
\end{equation}

Where $Q_p$ represents the product complexity derived from international trade data through the EFC method. The resulting quantification of territorial economic complexity, which computes a weighted sum over the local matrix $M$ using externally obtained complexity measures as weights, is referred to as \textit{Exogenous Fitness}. In our study, we further extend this framework by introducing a weighted-average formulation, in which the weighted sum is normalized by the total weight associated with each entity. To clearly distinguish between the two formulations, we refer to the original weighted-sum approach as \textit{Exogenous Fitness (sum)} and the weighted-average variant as \textit{Exogenous Fitness (average)}. Specifically, \textit{Exogenous Fitness (average)} differs from \textit{Exogenous Fitness (sum)} only in the definition of the intermediate quantity $\tilde{F}_s$. Instead of using a weighted sum, we compute a weighted average, i.e., $\tilde{F}_s = \displaystyle\frac{\sum_p M_{sp} Q_p}{\sum_p M_{sp}}$, while the normalization step remains unchanged. Additionally, we extend this approach by employing PCI-derived exogenous measures, leading to \textit{Exogenous ECI (sum)}. We further replace the weighted sum with a weighted average, resulting in \textit{Exogenous ECI (average)}.\\ When needed, we harmonize each country’s product codes to these HS6 measures (via code conversions and aggregation) so that globally benchmarked product metrics can be used to derive exogenous complexity values for individual countries. Because prior work did not address the aggregation of product codes, it offered no guidance on aggregation rules. In our implementation, we experiment with both summation and averaging schemes; the results show that the summation approach yields markedly stronger correlations with other indicators.

\subsubsection{Extensive Fitness}\label{subsec5}

Beyond preprocessing the matrix with RCA before applying the EFC algorithm, Tacchella et al.\cite{tacchella2012fitness} proposed using Market Share (MS) to process the data: $\text{MS}_{cp} = \frac{X_{cp}}{\sum_c X_{cp}}$; in this case, the algorithm uses a matrix which represents the importance of country $c$ in the global exports of product $p$ in terms of the market share (MS). The EFC computation based on MS-weighted data is referred to as the \textit{Extensive Fitness} in analogy with thermodynamics. Indeed, this measure is correlated with countries' size (in terms of total exports, GDP, or population) in contrast with the traditional measures, which are usually compared with GDP per capita and other "intensive" economic measures.
The extensive fitness method exhibits persistence and high predictive power with respect to RCA-based implementations, and is particularly suitable for higher levels of aggregations (i.e., when the total number of products is relatively low)~\cite{patelli2022MS}. The $\textbf{MS}$ matrix can also be used in combination with the country-level assessments of the complexity of products; in this case, we will call it the \textit{Exogenous Extensive Fitness}.

\subsubsection{Job Based Extensive Fitness}\label{subsec6}

This measure of economic complexity exploits the occupational structure of territories \cite{russo2025jobbased}. Following Aufiero et al. \cite{aufiero2024jobfitness}, one applies the Economic Fitness and Complexity (EFC) method to a matrix connecting jobs and skills. This allows us to derive $F_j$, the fitness of job $j$. Jobs with higher $F_j$ require a combination of more complex skills, whereas jobs with lower $F_j$ typically rely on simpler and more common skills. Building on the fitness of jobs, one can construct a job-based measure of economic complexity \cite{russo2025jobbased}, denoted as $Q^{JB}_i$. Specifically, the job-based complexity of industry $i$ is defined as the weighted average of the fitness of the jobs present in that industry:

\begin{equation}
Q_{i}^{JB} = \frac{\sum_j N_{ji} F_j}{\sum_j N_{ji}},
\end{equation}

where $N_{ji}$ denotes the number of employees in job $j$ within industry $i$. This measure captures the underlying, or hidden, complexity of industries.

Next, using a concordance table between product classifications, we map each industry $i$ to the corresponding product $p$ and the respective value of $Q_i^{JB}$. These values are then treated as exogenous complexity measures. By using market shares to construct the matrix \textbf{MS}, and applying the Exogenous Fitness (sum) method, we obtain the final measure of Job-Based Extensive Fitness (JBEF) at the territorial level:

\begin{equation}
\text{JBEF}_c = \sum_p MS_{cp} \, Q_p^{JB}.
\label{eq:JBEF}
\end{equation}


\section{Conclusions}\label{sec13}

A large body of literature studying economic development and innovation applies methods from complex networks and machine learning to investigate the so‑called economic complexity. The aim is to condense into a single index the capability content, diversification, and sophistication of an economic system; this is achieved by applying iterative algorithms to the matrix that assigns to each country the products it exports. The outputs of these algorithms are economic complexity metrics, among which the Economic Complexity Index (ECI) and Fitness are the most prominent. These measures have been successfully used to explain a wide range of heterogeneous phenomena, including inequality, economic growth, and $CO_2$ emissions.
Originally developed for analyses at the national level, these approaches have subsequently been extended to the subnational scale, where they have likewise provided an informative perspective on a variety of economic variables and development trajectories. The primary focus of this literature has consistently been on estimates of territorial complexity, even though these metrics inherently also produce, as an output, estimates of the economic complexity of products or industries that territories produce or export.\\
In this work, we start precisely from this underutilized output to assess the internal consistency of economic complexity metrics. Both \textit{global} measures—computed from international trade data—and \textit{local} measures—computed from subnational export or production data—provide estimates of the economic complexity of the same products. From a theoretical perspective, however, the capability content, technological sophistication, and degree of innovation embodied in a given product should not vary too much.
Instead, we find that product‑level complexity values computed at the global and local levels are not correlated with each other. We argue that this implies that at least one of the two estimates is fundamentally flawed. To determine which one, we compare both sets of measures with a simple and well‑established indicator of industrial sophistication: PRODY, defined as the weighted average of the GDP per capita of the countries exporting a given product. Only the product complexity computed from international trade data is correlated with PRODY.
Based on this evidence, we rule out the possibility of applying economic complexity algorithms \textit{endogenously} at the subnational level—that is, of running them directly on territory–product matrices.\\
This leads us to the following methodological choice for estimating the economic complexity of a territory: it should be computed \textit{exogenously}, by combining the territory–product matrix with product‑level complexity estimates calculated at the global level, without running any iterative algorithm. Even under this restriction, several options remain available, since, for a given territory, the complexities of the products it exports can be aggregated either by averaging or by summation, and with different weighting schemes.
To discriminate among these alternatives, we compare a set of candidate metrics with two fundamental economic indicators: GDP per capita, employment rate, and urbanization rate. We argue that, in order to effectively capture territorial capabilities, technological content, and innovative capacity, economic complexity measures must exhibit a positive association with such basic economic quantities. Moreover, this association should be \textit{consistent}, meaning not only strong but also approximately stable across different countries and across all geographical scales considered.
We find that the metrics satisfying these requirements are the \textit{extensive} measures, namely those based on the weighted sum of exogenous product complexities, where the weights are given by market shares—that is, by the relative contribution of each territorial unit to national exports.\\
\backmatter


\bmhead{Acknowledgements}

The authors would like to thank Stefano Boccaletti and Luciano Pietronero for interesting discussions.


\begin{appendices}




\end{appendices}



\end{document}